\documentclass[a4paper,11pt]{article}
\usepackage{latexsym}
\usepackage{amsmath, amsthm, amssymb, bbm}
\usepackage[mathscr]{eucal}
\usepackage{tikz}
\usepackage{enumerate}

\theoremstyle{plain}

\theoremstyle{definition}

\theoremstyle{remark}

\newcommand{\rhs}{r.h.s.\ }

\newcommand{\wrt}{w.r.t.\ }

\newcommand{\ud}{\mathrm{d}}
\newcommand{\del}{\partial}

\newcommand{\betrag}[1]{\lvert #1 \rvert}
\newcommand{\V}[1]{\mathbf{#1}}
\newcommand{\R}{\mathbb{R}}
\newcommand{\schw}{\mathcal{S}}
\newcommand{\skal}[2]{\langle #1 , #2 \rangle}

\DeclareMathOperator{\WF}{WF}
\DeclareMathOperator{\sd}{sd}
\newcommand{\N}{\mathbb{N}}
\newcommand{\Star}{ \, {\star} \, }

\begin{document}

\title{Ultraviolet-infrared mixing on the noncommutative Minkowski space in the Yang-Feldman formalism}
\author{Jochen Zahn \\ Courant Research Centre ``Higher Order Structures'' \\ University of G\"ottingen \\ Bunsenstra{\ss}e 3-5, D-37073 G\"ottingen, Germany}
\maketitle

\begin{abstract}
 We study infrared divergences due to ultraviolet-infrared mixing in quantum field theory on Moyal space with Lorentzian signature in the Yang-Feldman formalism. Concretely, we are considering the $\phi^4$ and the $\phi^3$ model in arbitrary even dimension. It turns out that the situation is worse than in the Euclidean setting, in the sense that we find infrared divergences in graphs that are finite there. We briefly discuss the problems one faces when trying to adapt the nonlocal counterterms that render the Euclidean model renormalizable.
\end{abstract}

\section{Introduction}

The most serious difficulty that shows up in the study of quantum field theories on Moyal space (NCQFT), cf.~\cite{Review} for a review, is a peculiar mixing of low and high energy scales, the so-called UV-IR mixing~\cite{MinwallaRaamsdonk}. In the Euclidean case, this leads to strange infrared divergences, which can only be renormalized with counterterms that either break translation invariance \cite{GW}, or are nonlocal \cite{Gurau}. Another, rather technical, difficulty is that the models on spaces with Euclidean and Lorentzian signature are not easily related in the case of space-time noncommutativity, i.e., if there is no timelike direction that commutes with all other directions.
In particular, a naive application in the Lorentzian case of the Feynman rules derived in the Euclidean, cf.~\cite{Filk}, leads to a violation of unitarity \cite{GomisMehen}. For the Lorentzian case, two different quantization procedures have been proposed, the Hamiltonian framework \cite{DFR} and the Yang-Feldman approach \cite{YF, BDFP02}. In general, contrary to the case of classical spacetime, these two approaches are inequivalent.

Despite of the unitarity problem and the missing correspondence with some theory in the Lorentzian sector, most of the work on NCQFT has been done in the Euclidean framework. In particular, the UV-IR mixing was found in that setting \cite{MinwallaRaamsdonk}. However, relatively little is known about the situation in the Lorentzian case. Since at least some of the momentum integrations in the Hamiltonian and the Yang-Feldman approach are restricted to the mass shell, there was hope that in these models the infrared divergences were absent or alleviated in the massive case, as it is the integration over the origin in momentum space that causes the troubles in the Euclidean. However, it was recently shown \cite{UVIRHamilton} that in the Hamiltonian approach a kind of UV-IR mixing occurs, even though the mechanism is quite different. Here, we show that also the Yang-Feldman approach is plagued by the UV-IR mixing.

Typically, the self-energy of a nonplanar (sub)graph will be a function of $(\sigma p)^2$, where $\sigma$ is the noncommutativity matrix, with a divergence in $(\sigma p)^2 = 0$. There are now two potential problems:
\begin{enumerate}[(i)]
 \item The integration over the singularity at $(\sigma p)^2 = 0$.
 \item The fact that $(\sigma p)^2$ does not fall off in some directions, i.e., the hypersurfaces $(\sigma p)^2 = \mathrm{const}$ are not compact, which may spoil integrability.
\end{enumerate}
While the first difficulty has some similarity with the Euclidean case (where $(\sigma p)^2 = 0$ means $p=0$), the latter difficulty is new.
It leads to divergences in situations that are finite in the Euclidean case. A particularly striking example of this is a divergence in the two-dimensional case.

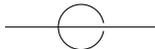
\begin{figure}[ht]
\begin{center}
\begin{tikzpicture}
 \draw (1,0) -- (3,0);
 \draw (1.7,0) arc (180:10:0.3);
 \draw (1.7,0) arc (-180:-10:0.3);
\end{tikzpicture}
\end{center}
\caption{The nonplanar tadpole.}
\label{fig:NonplanarTadpole}
\end{figure}

\begin{figure}[ht]
\begin{center}
\begin{tikzpicture}
 \draw (5,2) to (6,3);
 \draw (5,2) to (4,3);
 \draw (5,2) to [out=225,in=180] (1,0);
 \draw (5,2) to [out=315,in=0] (9,0);
 \draw (1,0) -- (3.5,0);
 \draw (7.5,0) -- (9,0);
 \draw (1.7,0) arc (180:10:0.3);
 \draw (1.7,0) arc (-180:-10:0.3);
 \draw (2.7,0) arc (180:10:0.3);
 \draw (2.7,0) arc (-180:-10:0.3);
 \draw (7.7,0) arc (180:10:0.3);
 \draw (7.7,0) arc (-180:-10:0.3);
 \draw (5.5,0) node {...};
\end{tikzpicture}
\end{center}
\caption{A graph exhibiting an infrared divergence due to UV-IR mixing.}
\label{fig:Phi4Euclidean}
\end{figure}
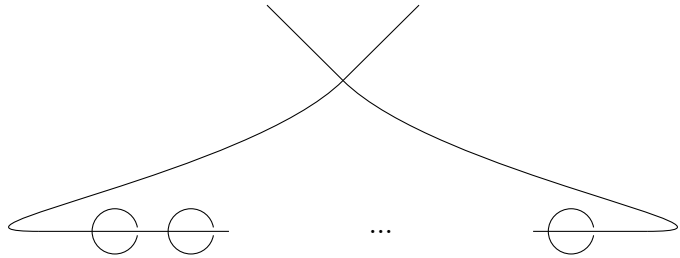

Let us briefly review how the UV-IR mixing occurs in the Euclidean setting. Some graphs that would be finite in the commutative case are regularized by the inverse of the incoming momentum. The simplest example is the so-called nonplanar tadpole in the $\phi^4$ model, which is depicted in the graph shown in Figure~\ref{fig:NonplanarTadpole}.
If such a graph is inserted in a loop of a bigger graph, then the momentum will be integrated over $p=0$, where it diverges. As an example, consider a graph of the form shown in Figure~\ref{fig:Phi4Euclidean}. Such a graph is always UV finite, as the nonplanar tadpoles fall off exponentially for large momentum $p$ (in the massive case). But the integral over the origin $p=0$ may lead to troubles.
In the two-dimensional case, the nonplanar tadpoles behave for small momenta as $\log p^2$, so an arbitrary number of them can be inserted without spoiling integrability at $p=0$. However,
in the four-dimensional case, the nonplanar tadpoles scale as $p^{-2}$. Thus, if two of them are inserted in the above graph, it diverges logarithmically in the infrared. With more and more nonplanar insertions, the behavior at $p = 0$ can be made arbitrarily bad.

The problem can be cured by modifying the propagator, either by introducing the so-called Grosse-Wulkenhaar potential \cite{GW}, which breaks translation invariance, or by adding a nonlocal but translation-invariant term \cite{Gurau}. However, an adaption of the Grosse-Wulkenhaar term to the Lorentzian case leads to strange divergences \cite{GWMink}, even in two dimensions. The setting of \cite{Gurau} has not yet been considered in the Lorentzian case, and we comment on that possibility below (in Section~\ref{sec:Conclusion}).

In the Yang-Feldman formalism, the loop momentum $p$ in graphs of the type depicted in Figure~\ref{fig:Phi4Euclidean} will always be on-shell. Thus, in a massive theory, one does not integrate over $p=0$. One might thus hope that in this framework the infrared divergences are absent or weakened. However, we show that the analog of the graph shown in Figure~\ref{fig:Phi4Euclidean} diverges in the Yang-Feldman formalism, already for one nonplanar insertion and independently of the dimension, i.e., even for $d=2$. Now a graph of this form is a tadpole, so one might think that the divergence can easily be subtracted, in particular as it is local in the adiabatic limit. However, we will see that the graph is finite before taking the adiabatic limit. Thus, a suitable counterterm can not be local, as it must depend on the cutoff function in a highly nontrivial way. We will also show that when nonplanar graphs are inserted into a planar fish graph, one finds the same dependence on the dimension as in the Euclidean case, i.e., a divergence in the $\phi^4_4$ and the $\phi^3_6$ model. However, the divergences appear earlier than in the Euclidean case, e.g., for the $\phi_6^3$ model, with two nonplanar insertions, as opposed to three in the Euclidean case.

The paper is organized as follows. In the next section, we give a short introduction to the Yang-Feldman formalism. In Section~\ref{sec:phi4}, we discuss the analog of the graph shown in Figure~\ref{fig:Phi4Euclidean} in the Yang-Feldman formalism. Section~\ref{sec:phi3} deals with the UV-IR mixing in the $\phi^3$ model. We conclude with a summary and an outlook. In an appendix, we recall some notions from microlocal analysis that are used below.

\subsection{Notation and conventions}
Throughout, we work on Moyal space with even dimension $d$. The $\Star$-product is defined via the twisted convolution of the Fourier transforms as
\begin{equation}
\label{eq:Star}
 (f \Star h) \hat{\ } (\tilde k) = (2\pi)^{-d/2} \int \ud^d k \ \hat{f}(k) \hat{h}(\tilde k - k) e^{-\frac{i}{2} k_\mu \sigma^{\mu \nu} \tilde k_\nu}.
\end{equation}
The noncommutativity matrix $\sigma$ is assumed to be given by\footnote{It is thus of the form proposed in \cite{DFR} in order to fulfill certain space-time uncertainty relations derived from semiclassical arguments. Note that we choose the length scale of noncommutativity as the length unit.}
\[
 \sigma = \begin{pmatrix} \epsilon &  & 0 \\  & \ddots &  \\ 0 &  & \epsilon  \end{pmatrix},
\]
with
\[ \epsilon = \begin{pmatrix} 0 & -1 \\ 1 & 0 \end{pmatrix}. \]
Thus, it is always the first spatial coordinate that does not commute with time. Correspondingly, we will often decompose $d$-dimensional vectors as $p = (p_0, p_1, p_s)$.
Given a $d-1$ dimensional momentum vector $\V{p}$, we define $\omega_p = \sqrt{\betrag{\V{p}}^2 + m^2}$ and $p^\pm = (\pm \omega_p, \V{p})$. As they play a major role in the Yang-Feldman formalism, we recall the retarded propagator and the two-point function in momentum space:
\begin{align*}
 \hat \Delta_R(k) & = (2\pi)^{-\frac{d}{2}} \lim_{\varepsilon \to +0} \frac{-1}{k^2-m^2+i\varepsilon k_0}, \\
 \hat \Delta_+(k) & = (2\pi)^{-\frac{d}{2}+1} \theta(k_0) \delta(k^2-m^2).
\end{align*}
As usual, $\mathcal{D}'(\R^n)$ denotes the distributions on test functions with compact support and $\schw'(\R^n)$ the tempered distributions. We sometimes use the notation $c_d$ for a constant that depends on the dimension and whose value may change in the same equation.

\section{The Yang-Feldman formalism}
We give a brief introduction to the Yang-Feldman formalism in the context of NCQFT. The basic idea is a perturbative and recursive construction of the interacting field in terms of the incoming field, which is supposed to be free. As an example, consider the noncommutative $\phi^4$ model. There, the equation of motion is given by
\begin{equation}
\label{eq:eom}
 (\Box + m^2) \phi = \lambda \phi \Star \phi \Star \phi.
\end{equation}
One now writes the interacting field as a formal power series in the coupling constant $\lambda$:
\[
 \phi = \sum_{n=0}^\infty \lambda^n \phi_n.
\]
Inserting this ansatz into \eqref{eq:eom}, one obtains
\begin{equation}
\label{eq:YFrecursion}
 (\Box + m^2) \phi_n = \sum_{\sum n_i = n-1} \phi_{n_1} \Star \phi_{n_2} \Star \phi_{n_3}.
\end{equation}
In particular, $\phi_0$ is a free field. Identifying it with the incoming field, the higher order components are obtained by convolution with the retarded propagator:
\begin{align}
 \phi_1 & = \Delta_R \times \phi_0 \Star \phi_0 \Star \phi_0, \nonumber \\
\label{eq:phi2_1}
 \phi_2 & = \Delta_R \times \left( \phi_1 \Star \phi_0 \star \phi_0 + \phi_0 \Star \phi_1 \star \phi_0 + \phi_0 \Star \phi_0 \star \phi_1 \right).
\end{align}
Quantum effects enter when contractions are considered. Two free fields $\phi_0$ can be contracted, yielding a two-point function $\Delta_+$.

A subtle point in the quantization procedure concerns a symmetry of the $\Star$-product. The change $\sigma \mapsto - \sigma$ corresponds to the replacement of the $\Star$-product by the $\bar \Star$-product, which is defined by $f \bar \Star g = g \Star f$. But, obviously, we have $\phi \Star \phi = \phi \bar \Star \phi$. Thus, the equation of motion \eqref{eq:eom} is invariant under $\sigma \mapsto - \sigma$. However, this symmetry is violated in a naive quantization\footnote{This observation is due to Micha{\l} Wrochna (private communication).}. As an example, consider the product $\phi \Star \phi \Star \phi$ occurring on the \rhs of \eqref{eq:eom}. By \eqref{eq:Star}, we would write it, in momentum space, as
\begin{multline}
\label{eq:phi3_wo_sym}
 \widehat{\phi \Star \phi \Star \phi}(k) = c_d \int \prod\nolimits_i \ud^dk_i \ \delta(k-\sum k_i) \hat \phi(k_1) \hat \phi(k_2) \hat \phi(k_3)  \\ \times e^{-\frac{i}{2} ( k_1 \sigma k_2 + k_1 \sigma k_3 + k_2 \sigma k_3 )}.
\end{multline}
If the $\hat \phi(k_i)$'s were numbers, this expression would be invariant under the replacement $\sigma \mapsto - \sigma$. However, in the quantum case, they are operators that do not commute in general. In order to restore the classical symmetry, we propose to symmetrize the quantum field part in the product that defines the interaction term, i.e., to set
\begin{multline}
\label{eq:phi3_w_sym}
\widehat{\phi \Star \phi \Star \phi}(k) = c_d \int \prod\nolimits_i \ud^dk_i \ \delta(k-\sum k_i) \{ \hat \phi(k_1), \hat \phi(k_2), \hat \phi(k_3) \} \\ \times e^{-\frac{i}{2} ( k_1 \sigma k_2 + k_1 \sigma k_3 + k_2 \sigma k_3 )}
\end{multline}
instead of \eqref{eq:phi3_wo_sym}, where $\{ \cdot, \cdot, \cdot \}$ stands for complete symmetrization\footnote{The invariance under the symmetry $\sigma \mapsto -\sigma$ could also be restored by just symmetrizing $\hat \phi(k_1)$ and $\hat \phi(k_3)$. However, we also have, e.g., $\phi \bar \Star (\phi \Star \phi) = (\phi \Star \phi) \Star \phi = \phi \Star (\phi \Star \phi)$. Thus, the classical expression is also invariant under the replacement of just one of the $\Star$-products by $\bar \Star$. In order to keep that symmetry, we use a complete symmetrization.}. Such a symmetrization was already proposed and used in \cite{NCQED} in order to cure certain inconsistencies in the quantization of gauge fields.
Using this product in \eqref{eq:YFrecursion}, we find
\begin{multline}
\label{eq:phin_w_sym}
 \hat \phi_n(k) = c_d \hat \Delta_R(k) \int \prod_i \ud^dk_i \ \delta(k-\sum k_i)  \sum_{\sum n_i=n-1} \hat \phi_{n_1}(k_1) \hat \phi_{n_2}(k_2) \hat \phi_{n_3}(k_3) \\ \times \left\{ \cos (\tfrac{1}{2} k_1 \sigma k_2) e^{-\frac{i}{2} ( k_1 + k_2) \sigma k_3} + \cos (\tfrac{1}{2} k_1 \sigma k_3) e^{-\frac{i}{2} ( k_1 + k_3) \sigma k_2} \right. \\ \left. + \cos (\tfrac{1}{2} k_2 \sigma k_3) e^{-\frac{i}{2} ( k_2 + k_3) \sigma k_1} \right\}.
\end{multline}
The formal power series thus obtained is then a solution of \eqref{eq:eom}, where the expression on the \rhs is defined by \eqref{eq:phi3_w_sym}.

\begin{figure}[ht]
\begin{center}
\begin{tikzpicture}[scale=0.5]
 \draw[double] (0,0) to (0,2);
 \draw (0,2) to (-1,4);
 \draw (0,2) to (1,4);
 \draw (0,2) to (0,4);
 \draw (0,0) node [below] {$k$};
 \draw (-1,4) node [above] {$k_1$};
 \draw (0,4) node [above] {$k_2$};
 \draw (1,4) node [above] {$k_3$};
 \draw (0,2) node [right] {$v(k_1, k_2, k_3)$};
\end{tikzpicture}
\end{center}
\caption{The graphical representation of $\phi_1$.}
\label{fig:phi4Vertex}
\end{figure}
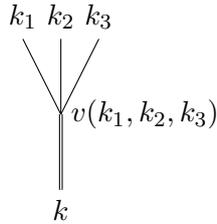

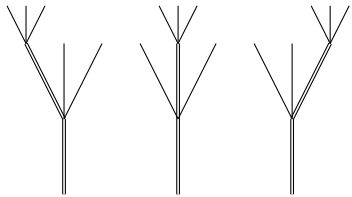
\begin{figure}[ht]
\begin{center}
\begin{tikzpicture}[scale=0.5]
 \draw[double] (0,0) to (0,2);
 \draw[double] (0,2) to (-1,4);
 \draw (0,2) to (1,4);
 \draw (0,2) to (0,4);
 \draw (-1,4) to (-1.5,5);
 \draw (-1,4) to (-1,5);
 \draw (-1,4) to (-0.5,5);
  \draw[double] (3,0) to (3,2);
  \draw[double] (3,2) to (3,4);
  \draw (3,2) to (4,4);
  \draw (3,2) to (2,4);
  \draw (3,4) to (2.5,5);
  \draw (3,4) to (3,5);
  \draw (3,4) to (3.5,5);
  \draw[double] (6,0) to (6,2);
  \draw[double] (6,2) to (7,4);
  \draw (6,2) to (5,4);
  \draw (6,2) to (6,4);
  \draw (7,4) to (6.5,5);
  \draw (7,4) to (7,5);
  \draw (7,4) to (7.5,5);
\end{tikzpicture}
\end{center}
\caption{The graphical representation of $\phi_2$.}
\label{fig:phi2}
\end{figure}

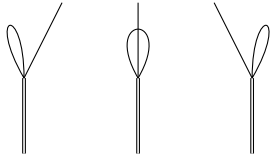
\begin{figure}[ht]
\begin{center}
\begin{tikzpicture}[scale=0.5]
 \draw[double] (0,0) to (0,2);
 \draw (0,2) to [controls=+(120:2) and +(90:2)] (0,2);
 \draw (0,2) to (1,4);
 \draw[double] (3,0) to (3,2);
 \draw (3,2) to [controls=+(120:2) and +(60:2)] (3,2);
 \draw (3,2) to (3,4);
 \draw[double] (6,0) to (6,2);
 \draw (6,2) to [controls=+(90:2) and +(60:2)] (6,2);
 \draw (6,2) to (5,4);
\end{tikzpicture}
\end{center}
\caption{The $\phi^4$ tadpole.}
\label{fig:tadpole}
\end{figure}

We now want to introduce a graphical notation.
The expression in curly brackets in \eqref{eq:phin_w_sym} defines the vertex factor. In the graphical notation, we would now express $\phi_1(k)$ by the graph shown in Figure~\ref{fig:phi4Vertex}.
In this notation, a double line stands for the retarded propagator. An open single line stands for an uncontracted free field $\phi_0$. A contraction is depicted by linking the two ends. Thus, a single line that links two (possibly coinciding) vertices stands for a two-point function. Higher order components of the field are obtained by replacing a single line by the same building block, cf. the recursive formula \eqref{eq:phin_w_sym}. As an example, the graphical representation of $\phi_2$ is depicted in Figure~\ref{fig:phi2}.
At first glance it does not seem to matter to which side the tree grows, i.e., the three graphs shown in Figure~\ref{fig:phi2} seem to be identical, as the vertex factor
\begin{multline*}
 v(k_1, k_2, k_3) = \cos (\tfrac{1}{2} k_1 \sigma k_2) e^{-\frac{i}{2} ( k_1 + k_2) \sigma k_3} + \cos (\tfrac{1}{2} k_1 \sigma k_3) e^{-\frac{i}{2} ( k_1 + k_3) \sigma k_2}  \\  + \cos (\tfrac{1}{2} k_2 \sigma k_3) e^{-\frac{i}{2} ( k_2 + k_3) \sigma k_1}
\end{multline*}
is invariant under permutations of the $k_i$'s. However, we recall that open single lines stand for free fields, and these do not necessarily commute. If one closes a loop by a contraction, one obtains the two-point function $\hat \Delta_+(k)$, where $k$ is the momentum going from left to right. But $\hat \Delta_+$ is not symmetric, so one has to take care about the order.
Another downside of our graphical notation is that we can not distinguish between planar and nonplanar graphs in this notation. To do that, one has to translate the graph to an analytic expression. As an example, consider the tadpole, i.e., the graphs shown in Figure~\ref{fig:tadpole} (in this case, the three different graphs yield the same result).
It is the contracted part of $\phi_1$, for which we obtain
\begin{align}
 \hat \phi_1^c(k) & = c_d \hat \Delta_R(k) \hat \phi_0(k) \int \ud^d p \ \hat \Delta_+(p) \left\{ 2 + \cos p \sigma k \right\} \nonumber \\
\label{eq:phi4tadpole}
 & = c_d \hat \Delta_R(k) \hat \phi_0(k) \left\{ 2 \Delta_+(0) + \Delta_1(\sigma k) \right\}.
\end{align}
Here we used
\[
 \Delta_1(x) = \tfrac{1}{2} \left( \Delta_+(x) + \Delta_+(-x) \right).
\]
The first term in \eqref{eq:phi4tadpole} diverges and corresponds to the usual tadpole. We subtract it by normal ordering. The second term, however is finite and nonlocal, so we do not subtract it (this procedure corresponds to the quasiplanar Wick products introduced in \cite{Quasiplanar}). This second term will be called the nonplanar tadpole in the following. In the Euclidean, one would find $\Delta_E(\sigma p)$ instead of $\Delta_1(\sigma p)$, where $\Delta_E$ is the Euclidean Green's function.

\section{The case of $\phi^4$}
\label{sec:phi4}

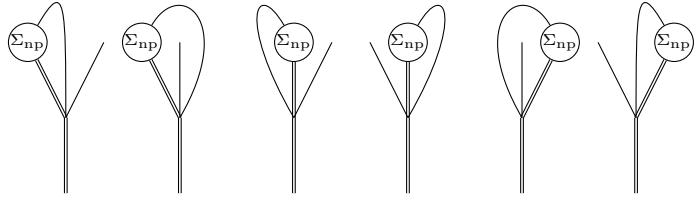
\begin{figure}[ht]
\begin{center}
\begin{tikzpicture}[scale=0.5]
 \draw[double] (0,0) to (0,2);
 \node [circle,inner sep=0pt,draw] (s1) at (-1,4) {{\tiny $\Sigma_\mathrm{np}$}};
 \draw[double] (0,2) to (s1);
 \draw (0,2) to (1,4);
 \draw (s1) to [controls=+(60:2) and +(90:3)] (0,2);
 \draw[double] (3,0) to (3,2);
 \node [circle,inner sep=0pt,draw] (s2) at (2,4) {{\tiny $\Sigma_\mathrm{np}$}};
 \draw[double] (3,2) to (s2);
 \draw (3,2) to (3,4);
 \draw (s2) to [controls=+(60:2) and +(60:3)] (3,2);
 \draw[double] (6,0) to (6,2);
 \node [circle,inner sep=0pt,draw] (s3) at (6,4) {{\tiny $\Sigma_\mathrm{np}$}};
 \draw[double] (6,2) to (s3);
 \draw (6,2) to (7,4);
 \draw (s3) to [controls=+(120:2) and +(120:3)] (6,2);
 \draw[double] (9,0) to (9,2);
 \node [circle,inner sep=0pt,draw] (s4) at (9,4) {{\tiny $\Sigma_\mathrm{np}$}};
 \draw[double] (9,2) to (s4);
 \draw (9,2) to (8,4);
 \draw (s4) to [controls=+(60:2) and +(60:3)] (9,2);
 \draw[double] (12,0) to (12,2);
 \node [circle,inner sep=0pt,draw] (s5) at (13,4) {{\tiny $\Sigma_\mathrm{np}$}};
 \draw[double] (12,2) to (s5);
 \draw (12,2) to (12,4);
 \draw (s5) to [controls=+(120:2) and +(120:3)] (12,2);
 \draw[double] (15,0) to (15,2);
 \node [circle,inner sep=0pt,draw] (s6) at (16,4) {{\tiny $\Sigma_\mathrm{np}$}};
 \draw[double] (15,2) to (s6);
 \draw (15,2) to (14,4);
 \draw (s6) to [controls=+(120:2) and +(90:3)] (15,2);
\end{tikzpicture}
\end{center}
\caption{The Yang-Feldman snowman graphs.}
\label{fig:snowman}
\end{figure}

We consider the snowman graphs of $\phi^4$, i.e., the tadpole with one inserted tadpole, which is part of the contracted part $\phi_2^c$ of $\phi_2$. As we subtracted the usual local divergence of the tadpole, only the nonplanar contribution, i.e., the second term in \eqref{eq:phi4tadpole}, remains. Thus, we have the self-energy
\[
 \Sigma_\mathrm{np}(k) = \Delta_1(\sigma k).
\]
The snowman graphs of $\phi^4$ are now depicted in Figure~\ref{fig:snowman}.
Here the encircled $\Sigma_\mathrm{np}$ stands for the nonplanar part of the tadpole self-energy. Thus, we obtain
\begin{equation}
\label{eq:phi2_no_cutoff}
 \hat \phi_2^c(k) = c_d \hat \Delta_R(k) \hat \phi_0(k) \int \ud^d p \ \hat \Delta_R(p) \Sigma_\mathrm{np}(p) \hat \Delta_1(p) \left\{ 2 + \cos p \sigma k \right\}.
\end{equation}

In order to separate the UV and the IR problem, we introduce an IR cutoff. We do this rather ad hoc by replacing momentum conservation at each vertex in the graphs in Figure~\ref{fig:snowman} by $\hat g(\sum k_i)$, where $\hat g$ is the Fourier transform of a smooth function $g$ with compact support. Later, we will consider the adiabatic limit in which this test function is replaced by a constant. With such a cutoff, we have, instead of \eqref{eq:phi2_no_cutoff},
\begin{multline}
\label{eq:phi2_cutoff}
 \hat \phi_2^c(k) = c_d \hat \Delta_R(k) \int \prod\nolimits_i \ud^d k_i \ \hat g(k-k_1-k_2-k_3) \hat \phi_0(k_1) \\ \times \hat \Delta_R(k_2) \hat g(k_2 + k_3) \Sigma_\mathrm{np}(k_3) \hat \Delta_1(k_3) \left\{ 2 + \cos k_2 \sigma k \right\}.
\end{multline}
We note that there is some ambiguity in giving a momentum to $\Sigma_\mathrm{np}$ and the second term in curly brackets, but the result in the adiabatic limit does not depend on this choice. The second term in curly brackets corresponds, in Euclidean NCQFT, to a nonplanar tadpole inserted in a nonplanar tadpole. It turns out to be finite, as we will show below. Thus, we focus on the first term in curly brackets in \eqref{eq:phi2_cutoff}, which in the Euclidean setting corresponds to the graph shown in Figure~\ref{fig:Phi4Euclidean} with one nonplanar insertion. Transformation to position space leads to
\begin{equation}
\label{eq:phi2}
 \phi^\mathrm{pl}_2(z) = c_d \int \ud^dy \ \Delta_R(z-y) g(y) \phi_0(y) \int \ud^dx \ \Delta_R(y-x) u(x-y) g(x),
\end{equation}
where $u$ is the inverse Fourier transform of
\[
 \hat u(p) = \hat \Delta_1(p) \Sigma_\mathrm{np}(p) = \hat \Delta_1(p) \Delta_1(\sigma p).
\]

In the following, we will study how the second integral behaves, in particular in the adiabatic limit. As a first step, we want to establish that the inverse Fourier transform $u$ of $\hat{u}$ really exists. We will then discuss whether its product with $\Delta_R$ is well-defined, and finally consider the integral in the adiabatic limit. As problems only show up in the adiabatic limit, we conclude that we are dealing with an infrared divergence which can not be renormalized with the usual local counterterms.

\subsection{The distribution $u$}

As is easily checked, $\sigma p$ is spacelike if $p$ is timelike. As $\hat \Delta_+(p)$ has singular support on $p^2 = m^2$ and $\Delta_+(x)$ on $x^2 = 0$, the singular supports of $\hat \Delta_1(p)$ and $\Delta_1(\sigma p)$ do not overlap. Thus, their product $\hat u$ is well-defined as an element of $\mathcal{D}'(\R^d)$, by H\"ormanders criterion, cf. Appendix~\ref{app:microlocal}. But it is not necessarily an element of $\schw'(\R^d)$. Hence, it is not clear whether its Fourier transform $u$ really exists. However, as $\hat \Delta_1(p)$ is supported on $\{ p | p^2 = m^2 \}$, also $\hat u(p)$ will be supported on this set. Furthermore, $\Delta_1(x)$ depends only on $x^2$, is singular on $\{ x | x^2 = 0 \}$, and is polynomially bounded (with all its derivatives) on $\{ x | \betrag{x^2} > \epsilon \}$ for any $\epsilon > 0$ . We have
\[
  (\sigma p)^2 = p_1^2 - p_0^2 - p_s^2 = - p^2 - 2 p_s^2,
\]
so that for $p^2 = m^2$ we have $(\sigma p)^2 \leq - m^2$. Thus, on the support of $\hat \Delta_1(p)$, $\Delta_1(\sigma p)$ is smooth and polynomially bounded (with all its derivatives). It follows that $\hat u$ is tempered, as $\hat \Delta_1$ is tempered and for $f \in \schw(\R^d)$ we may define $\skal{\hat u}{f} = \skal{\hat \Delta_1}{\psi f}$, where $\psi$ is smooth, polynomially bounded (with all its derivatives), and coincides with $\Delta_1(\sigma \cdot)$ in a neighborhood of the support of $\hat \Delta_1$.

As $\hat u \in \schw'(\R^d)$, its Fourier transform $u$ is well-defined. If either the support of $\Delta_1$ or that of $\hat \Delta_1$ were compact, then from $u = \Delta_1 \times \hat \Delta_1(\sigma^{-1} \cdot)$ and \cite[Thm.~8.2.14]{Hoermander} we could conclude that the wave front set, cf. Appendix~\ref{app:microlocal}, of $u$ is contained in 
\[
  \{ (x, k) | \exists y \text{ s.t. } (y,k) \in \WF(\Delta_1), (x-y,k) \in \WF(\hat \Delta_1(\sigma ^{-1} \cdot)) \}.
\]
As the cotangent vectors of $\WF(\hat \Delta_+(\sigma ^{-1} \cdot))$ always point in spacelike directions and those of $\WF(\Delta_+)$ in lightlike directions, cf. Appendix~\ref{app:microlocal}, this would imply $\WF(u)  = \emptyset$. But as neither of the two distributions has compact support, the singular support may be enlarged by infrared divergences, as we will see now. We write $\Delta_1(x) = h(x^2)$, where $h(y)$ is smooth apart from $y=0$ and falls off exponentially for $y \to - \infty$.
We then formally compute
\begin{align}
 u(x) & = c_d \int \ud^dp \ \hat \Delta_1(p) \Delta_1(\sigma p) e^{-i x p} \nonumber\\
\label{eq:u_integral}
 & = c_d \int \frac{\ud^{d-1}\V{p}}{2\omega_{p}} h(-2 p_s^2 - m^2) \cos (x p^+).
\end{align}
It is tempting to interpret this as an oscillatory integral, cf.~\cite{RS2}, but this is not possible, as derivatives \wrt $p_s$ do not lower the degree of the would-be symbol. Instead, we first carry out the $p_1$ integration:
\begin{align*}
 u(x) & = c_d \int \ud^{d-2}p_s \ h(-2 p_s^2 - m^2) e^{i x_s \cdot p_s} \\
  & \quad \times \int_{-\infty}^\infty \ud p_1 \frac{\cos (x^0 \sqrt{p_1^2 + p_s^2 + m^2} - x^1 p_1)}{2 \sqrt{p_1^2 + p_s^2 + m^2}}  \\
  & = c_d \int \ud^{d-2}p_s \ h(-2 p_s^2 - m^2) e^{i x_s \cdot p_s} \Delta_1^{(2)}(x^0, x^1; \sqrt{p_s^2 + m^2}).
\end{align*}
Here $\Delta_1^{(2)}(x;m)$ denotes $\Delta_1(x)$ in two dimensions for mass $m$. It has singular support on the light cone, where it diverges logarithmically. Away from the singularity, it is bounded as a function of $m$ for $m \to \infty$. Thus, the remaining integral over $p_s$ is well defined, yielding a distribution with singular support in $x^0 = \pm x^1$, where it diverges logarithmically. We emphasize that this divergence is independent on the dimension $d$.

Having established that $\hat u$ is tempered, and thus that $u$ exists, we may now discuss whether the point-wise product $\Delta_R(x) u(-x)$ appearing in \eqref{eq:phi2} is well-defined. As both distributions are singular at the origin and the cotangent component of the wave front set of $\Delta_R$ points in every direction at that point, their product is not defined in the sense of H\"ormander. However, we established that $u$ diverges logarithmically at the origin, it thus has scaling degree $0$ there, cf. Appendix~\ref{app:microlocal}. As the scaling degree of $\Delta_R$ is $d-2$ at the origin, their product has scaling degree $d-2$ and is thus unambiguously extendable to the origin. The same argument also applies to the one-dimensional submanifold $\{x^0 = \pm \betrag{x^1}, x_s = 0\}$ on which both distributions are singular. It follows that the integral over $x$ in \eqref{eq:phi2} is well-defined, as long as $g$ is a test function. However, as we will see below, problems appear in the adiabatic limit, where $g$ is replaced by a constant.

\subsection{The adiabatic limit}
In the previous subsection, we established that the integral over $x$ in \eqref{eq:phi2} is indeed well-defined as long as $g$ is a test function. In the adiabatic limit, we formally obtain
\begin{equation*}
 \phi^\mathrm{pl}_2(z) = c_d \int \ud^dy \ \Delta_{R}(z-y) \phi_0(y) \int \ud^dx \ \Delta_R(-x) u(x).
\end{equation*}
The integral over $x$ is a formal integral, which we abbreviate by $\Pi$.
As it is formal anyway, we feel free to apply formal Fourier transformation and obtain
\begin{align*}
 \Pi & = c_d \widehat{\Delta_A u}(0) \\
     & = c_d (\hat \Delta_A \times \hat u) (0) \\  
     & = c_d \int \ud^dp \ \Delta_1(\sigma p) \hat \Delta_1(p) \hat \Delta_R(p) \\
     & = c_d \int \ud^dp \ \Delta_1(\sigma p) \hat \Delta_+(p) \left( \hat \Delta_R(p) + \hat \Delta_A(p) \right).
\end{align*}
In the last step, we used that $\Delta_1$ is symmetric.
Now formally (and rigorously in an adiabatic limit, cf. \cite{AdLim}) we have
\[
 \hat \Delta_+(p) \left( \hat \Delta_R(p) + \hat \Delta_A(p) \right) = c_d \del_{m^2} \hat \Delta_+(p).
\]
Thus, we obtain
\[
 \Pi = c_d \int \ud^dp \ \Delta_1(\sigma p) \del_{m^2} \hat \Delta_+(p).
\]
We have
\[
 \int \ud^dp \ f(p) \del_{m^2} \hat \Delta_+(p) = c_d \int \ud^{d-1} \V{p} \left( \frac{1}{4 \omega_\V{p}^3} f(p^+) - \frac{1}{4 \omega_\V{p}^2} \del_0 f(p^+) \right).
\]
Hence, the above yields
\[
 \Pi = c_d \int \ud^{d-1} \V{p} \left( \frac{1}{4 \omega_\V{p}^3} h((\sigma p^+)^2) + \frac{1}{2 \omega_\V{p}} h'((\sigma p^+)^2) \right),
\]
where we introduced again the notation $\Delta_1(x) = h(x^2)$. By the same argument as in the previous subsection, the second term of this integral diverges logarithmically. As discussed above, it should be termed an IR divergence. We emphasize that the divergence shows up in any dimension, in particular also for $d=2$. In the Euclidean framework, the problem was present only for $d \geq 4$. Furthermore, in the case $d=4$, two nonplanar tadpoles had to be introduced into the loop to see the divergence. In this sense, the Lorentz structure deteriorates the situation. This divergence is an instance of the difficulty (ii) mentioned in the introduction.

\begin{figure}[ht]
\begin{center}
\begin{tikzpicture}
 \draw (5,2) to (5,1);
 \draw (5,1.5) to [out=180,in=180] (1,0);
 \draw (5,1.5) to [out=0,in=0] (9,0);
 \draw (1,0) -- (3.5,0);
 \draw (7.5,0) -- (9,0);
 \draw (1.7,0) arc (180:10:0.3);
 \draw (1.7,0) arc (-180:-10:0.3);
 \draw (2.7,0) arc (180:10:0.3);
 \draw (2.7,0) arc (-180:-10:0.3);
 \draw (7.7,0) arc (180:10:0.3);
 \draw (7.7,0) arc (-180:-10:0.3);
 \draw (5.5,0) node {...};
\end{tikzpicture}
\end{center}
\caption{The nonplanar tadpole with nonplanar insertions.}
\label{fig:phi4_np_np}
\end{figure}
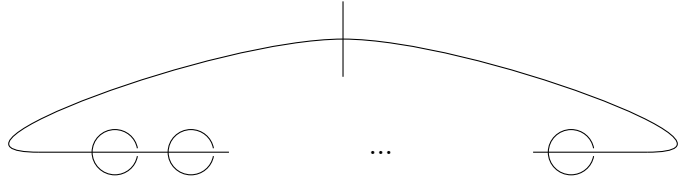

It remains to discuss the second term in curly brackets in \eqref{eq:phi2_cutoff}, which we ignored up to now. In the adiabatic limit, it is given by
\[
 \phi_2^\mathrm{np}(k) = c_d \hat \Delta_R(k) \hat \phi_0(k) \int \ud^4 p \ \del_{m^2} \hat \Delta_+(p) h((\sigma p)^2) \cos p \sigma k,
\]
with $h$ as above. This is very similar to the expression for $u(x)$ given in \eqref{eq:u_integral}, the difference beeing that $\hat \Delta_+$ is replaced by $\del_{m^2} \hat \Delta_+$ and $x$ by $\sigma k$. As above, the first replacement does not change the asymptotic behavior. However, the presence of $\hat \phi_0(k)$ forces $k$ to the mass shell, so $\sigma k$ is spacelike. By the argument given in the previous subsection, the integral over $p$ is then finite. Thus, the nonplanar tadpole with inserted nonplanar tadpole is finite. This argument even goes through for an arbitrary number of nonplanar insertions. This is again in contrast to the Euclidean case, where the nonplanar tadpole with nonplanar insertions, i.e., the graph shown in Figure~\ref{fig:phi4_np_np}, has the same infrared problems as the graph shown in Figure~\ref{fig:Phi4Euclidean}.

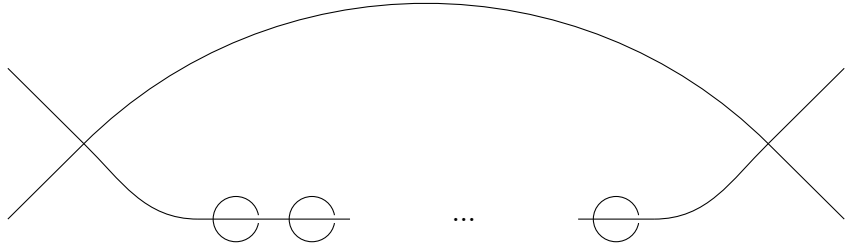
\begin{figure}[ht]
\begin{center}
\begin{tikzpicture}
 \draw (0,1) to (1,0);
 \draw (0,-1) to (1,0);
 \draw (10,0) to (11,1);
 \draw (10,0) to (11,-1);
 \draw (1,0) to [out=45,in=135] (10,0);
 \draw (1,0) to [out=-45,in=180] (2.5,-1);
 \draw (2.5,-1) -- (4.5,-1);
 \draw (7.5,-1) -- (8.5,-1);
 \draw (2.7,-1) arc (180:10:0.3);
 \draw (2.7,-1) arc (-180:-10:0.3);
 \draw (3.7,-1) arc (180:10:0.3);
 \draw (3.7,-1) arc (-180:-10:0.3);
 \draw (7.7,-1) arc (180:10:0.3);
 \draw (7.7,-1) arc (-180:-10:0.3);
 \draw (8.5,-1) [out=0,in=225] to (10,0);
 \draw (6,-1) node {...};
\end{tikzpicture}
\end{center}
\caption{Another example of a $\phi^4$ with infrared divergence due to UV-IR mixing in the Euclidean setting.}
\label{fig:phi4fish}
\end{figure}

Finally, we consider what happens when the nonplanar tadpoles are inserted into a fish graph, i.e., in the Euclidean setting, a graph of the form shown in Figure~\ref{fig:phi4fish}.
The situation is then analogous to the situation discussed in the next section, i.e., of the $\phi^3$ model. As is shown there, such a graph is divergent for two nonplanar insertions if the nonplanar subgraph behaves for small $(\sigma p)^2$ as $\Sigma_\mathrm{np}(p) \sim (\sigma p)^{-2}$ or worse. This is the case for $d \geq 4$. 

\section{The case of $\phi^3$}
\label{sec:phi3}
\begin{figure}[ht]
\begin{center}
\begin{tikzpicture}
 \draw (0,0) to (1,0);
 \draw (10,0) to (11,0);
 \draw (1,0) to [out=45,in=135] (10,0);
 \draw (1,0) to [out=-45,in=180] (2,-1);
 \draw (2,-1) to [controls=+(60:1.5) and +(240:1.5)] (3.5,-1);
 \draw (2,-1) to [controls=+(300:1.5) and +(120:1.5),draw=white,double=black,double distance=0.4pt,thick] (3.5,-1);
 \draw (3.5,-1) to (4,-1);
 \draw (4,-1) to [controls=+(60:1.5) and +(240:1.5)] (5.5,-1);
 \draw (4,-1) to [controls=+(300:1.5) and +(120:1.5),draw=white,double=black,double distance=0.4pt,thick] (5.5,-1);
 \draw (5.5,-1) to (5.7,-1);
 \draw (7.3,-1) to (7.5,-1);
 \draw (7.5,-1) to [controls=+(60:1.5) and +(240:1.5)] (9,-1);
 \draw (7.5,-1) to [controls=+(300:1.5) and +(120:1.5),draw=white,double=black,double distance=0.4pt,thick] (9,-1);
 \draw (9,-1) [out=0,in=225] to (10,0);
 \draw (6.5,-1) node {...};
\end{tikzpicture}
\end{center}
\caption{A $\phi^3$ graph exhibiting UV-IR mixing in the Euclidean setting.}
\label{fig:phi3Euclidean}
\end{figure}
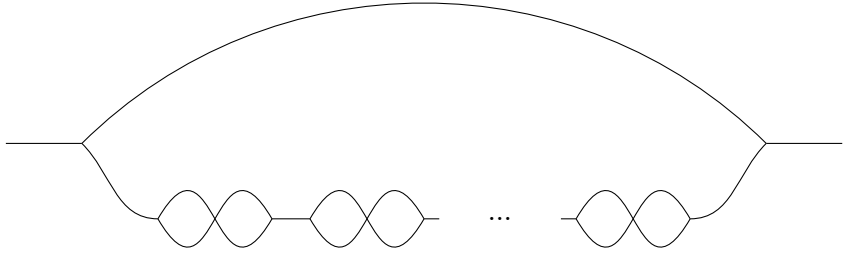

We now consider the case of the $\phi^3$ model. In the Euclidean case, the UV-IR mixing then occurs in graphs of the form shown in Figure~\ref{fig:phi3Euclidean}.
The fish graph in six dimensions is quadratically divergent, so in the Euclidean setting this translates into a scaling $p^{-2}$ for small momenta $p$ in the nonplanar fish graph. If three such nonplanar fish graphs are considered in a row, this gives a $p^{-6}$ scaling, which yields a logarithmic infrared divergence.

\subsection{The fish graph}
\begin{figure}[ht]
\begin{center}
\begin{tikzpicture}[scale=0.5]
 \draw (0,0) [double] to (0,2);
 \draw (0,2) [double] to (-1,4);
 \draw (-1,4) to (-2,6);
 \draw (-1,4) to [controls=+(60:2) and +(60:2)] (0,2);
 \draw (3,0) [double] to (3,2);
 \draw (3,2) [double] to (4,4);
 \draw (4,4) to (5,6);
 \draw (4,4) to [controls=+(120:2) and +(120:2)] (3,2);
\end{tikzpicture}
\end{center}
\caption{The $\phi^3$ fish graphs in the Yang-Feldman formalism.}
\label{fig:phi3fish}
\end{figure}
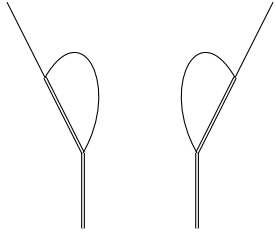

We consider a single fish graph in the Yang-Feldman formalism in arbitrary dimension, i.e., the graphs shown in Figure~\ref{fig:phi3fish}.
The graphs obtained by letting the uncontracted free field leave the upper vertex to the other side, yield the same result because of the symmetry of the vertex factor
\[
 v(k_1, k_2) = \cos \tfrac{1}{2} k_1 \sigma k_2.
\]
For the self-energy corresponding to these graphs, one thus finds
\[
 \Sigma(p) = c_d \int \ud^dk \ \hat \Delta_1(k) \hat \Delta_R(p-k) \left\{ 1 + \cos (k \sigma p) \right\}.
\]
The first term in curly brackets is the usual commutative $\phi^3$ fish graph. It is divergent for $d \geq 4$ and has to be renormalized by mass and possibly (depending on the dimension) field strength counterterms. The second term in curly brackets corresponds to the nonplanar fish graph known from the Euclidean theory. For timelike outer momentum $0 < p^2 < 4m^2$ and $d=4$, it was rigorously defined in the sense of oscillatory integrals, cf.~\cite{RS2}, in \cite{NCDispRel}. In this way, it could be reduced to a one-dimensional absolutely convergent integral. This can easily be generalized to arbitrary dimension, yielding
\begin{align*}
 \Sigma_\mathrm{np}(p) & = c_d \int \ud^dk \ \hat \Delta_+(k) \left(\hat \Delta_R(p-k) + \hat \Delta_R(p+k) \right) \cos (k \sigma p) \\
& = c_d \int_0^\infty \ud k \frac{k^{d-2}}{\omega_k (p^2 - 4 \omega_k^2)} \frac{\sin ( k \sqrt{\betrag{(\sigma p)^2}} )}{k \sqrt{\betrag{(\sigma p)^2}}}.
\end{align*}
For $d>4$, this is no longer absolutely convergent, but still defined as an oscillatory integral. In the limit $(\sigma p)^2 \to 0$, we find a logarithmic divergence $\Sigma_\mathrm{np}(p) \sim \log \betrag{(\sigma p)^2}$ for $d=4$, and a quadratic divergence $\Sigma_\mathrm{np}(p) \sim \betrag{(\sigma p)^2}^{-1}$ for $d=6$.

In the following, it is rather the behavior for spacelike outer momentum that is important. In that case, the loop integral can not be defined as an oscillatory integral. However, a formal calculation is feasible. We consider the case where $p$ is spacelike and $y=\sigma p$ timelike. Under the assumption that the self-energy of the nonplanar fish graph is well-defined, it is a function only of $p^2$ and $y^2$, by Lorentz invariance. By a Lorentz transformation, we can achieve $y = (y, \V{0})$. As $p$ and $y$ are orthogonal, we then have $p = (0, \V{p})$. For the self-energy, we thus obtain
\begin{align*}
 \Sigma_\mathrm{np}(p) & = c_d \int \ud^dk \ \hat \Delta_+(k) \left(\hat \Delta_R(p-k) + \hat \Delta_R(p+k) \right) \cos (k \cdot y) \\
& = c_d \int \frac{\ud^{d-1} \V{k}}{2\omega_k} \left( \frac{-1}{p^2 + 2 \V{p} \cdot \V{k} - i \varepsilon} + \frac{-1}{p^2 - 2 \V{p} \cdot \V{k} + i \varepsilon} \right) \cos (\omega_k y).
\end{align*}
For $d=2$ this integral is absolutely convergent, independently of $y$. For $d>2$, we carry out the integration over all but the azimuthal angle and obtain
\begin{align}
 \Sigma_\mathrm{np}(p) & = c_d \int_0^\infty \ud k \frac{k^{d-2}}{2\omega_k} \cos (\omega_k y) \nonumber \\ 
 & \quad \quad \quad \times \int_{-1}^1 \ud x \left( \frac{-1}{p^2 + 2 \sqrt{\betrag{p^2}} k x - i \varepsilon} + \frac{-1}{p^2 - 2 \sqrt{\betrag{p^2}} k x + i \varepsilon} \right) \nonumber \\
\label{eq:Sigma_phi3}
 & = c_d \int_0^\infty \ud k \frac{k^{d-2}}{2\omega_k} \frac{1}{2 \sqrt{\betrag{p^2}} k} \log \frac{(2 \sqrt{\betrag{p^2}} k - p^2)^2 + \varepsilon^2}{(2 \sqrt{\betrag{p^2}} k + p^2)^2 + \varepsilon^2} \cos (\omega_k y).
\end{align}
For nonvanishing $\varepsilon$, the integrand is smooth. Interpreting the cosine as the phase function, the remainder of the integrand is a symbol of order $d-5$, as
\[
 \frac{(2 \sqrt{\betrag{p^2}} k - p^2)^2 + \varepsilon^2}{(2 \sqrt{\betrag{p^2}} k + p^2)^2 + \varepsilon^2} \simeq 1 + \frac{2 \sqrt{\betrag{p^2}}}{k} \ \text{ for large } k.
\]
Thus, the integral \eqref{eq:Sigma_phi3} is well defined as an oscillatory integral, but diverges for small $y$ as $\log y$ for $d=4$ and as $y^{-2}$ for $d=6$. Hence, also for spacelike $p$ and timelike $\sigma p$, we find a logarithmic divergence $\Sigma_\mathrm{np}(p) \sim \log \betrag{(\sigma p)^2}$ for $d=4$, and a quadratic divergence $\Sigma_\mathrm{np}(p) \sim \betrag{(\sigma p)^2}^{-1}$ for $d=6$. We thus see the same scaling behavior as in the Euclidean setting, with the difference that in the present case we have a singularity not only at $p=0$ but on the hypersurface $(\sigma p)^2 = 0$.

As we could not define $\Sigma_\mathrm{np}(p)$ as an oscillatory integral on the whole $\R^d$, it is not clear that it defines a distribution on $\R^d$. In particular, it is not clear whether the singularity in $(\sigma p)^2 = 0$ is regularized by some $i \varepsilon$ or principal value description. In the following, we assume that this is the case and that its wave front set coincides with that of the nonplanar tadpole ($\Sigma_\mathrm{np}(p) = \Delta_1(\sigma p)$), i.e.,
\begin{equation}
\label{eq:wfSigma}
 \WF(\Sigma_\mathrm{np}) = \{ (k,y) | (\sigma k)^2 = 0, y = \lambda k, \lambda \neq 0 \}.
\end{equation}
Furthermore, we assume that, as for the nonplanar tadpole, $\Sigma_\mathrm{np}(p)$ falls off exponentially as $(\sigma p)^2 \to - \infty$. In our opinion, these are the most optimistic assumptions one can reasonably make. But even with these, one finds divergences, even some that are absent in the Euclidean case.

\subsection{Infrared divergences}

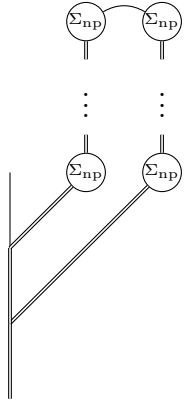
\begin{figure}[ht]
\begin{center}
\begin{tikzpicture}[scale=0.5]
 \node [circle,inner sep=0pt,draw] (s1) at (4,6) {{\tiny $\Sigma_\mathrm{np}$}};
 \node [circle,inner sep=0pt,draw] (s2) at (4,10) {{\tiny $\Sigma_\mathrm{np}$}};
 \node [circle,inner sep=0pt,draw] (s3) at (2,6) {{\tiny $\Sigma_\mathrm{np}$}};
 \node [circle,inner sep=0pt,draw] (s4) at (2,10) {{\tiny $\Sigma_\mathrm{np}$}};
 \draw[double] (0,0) to (0,2);
 \draw[double] (0,2) to (s1);
 \draw[double] (0,2) to (0,4);
 \draw[double] (s1) to (4,7);
 \draw (4,8) node {$\vdots$};
 \draw[double] (4,9) to (s2);
 \draw (s4) to [bend left] (s2);
 \draw (0,4) to (0,6);
 \draw[double] (0,4) to (s3);
 \draw[double] (s3) to (2,7);
 \draw (2,8) node {$\vdots$};
 \draw[double] (2,9) to (s4);
\end{tikzpicture}
\end{center}
 \caption{A $\phi^3$ graph in the Yang-Feldman formalism that does not suffer from UV-IR mixing.}
 \label{fig:phi3YF1}
\end{figure}

In the Yang-Feldman formalism, there are two possibilities
to set up graphs similar to the one shown in Figure~\ref{fig:phi3Euclidean}. The point is that the subgraphs consisting just of retarded propagators are always trees. Thus, a loop can only be closed by a two-point function. This, in turn, can be done either in the branch without insertions or in the one with insertions. The graph shown in Figure~\ref{fig:phi3YF1} is an example for the latter case.
In such a graph, the momentum in the branch containing the insertions is confined to the mass shell, so that one does not integrate over the singularity in $(\sigma p)^2 = 0$ (including the point $p=0$ that causes the trouble in the Euclidean). Also the integration along the direction $p_1$, that leads to the divergence in the $\phi^4$ model discussed in the previous section, is not problematic, as the retarded propagator in the other branch contributes another factor of $\frac{1}{\omega}$. As we assumed that $\Sigma_\mathrm{np}(p)$ falls off exponentially for $(\sigma p)^2 \to - \infty$, i.e., for $p_s \to \infty$, the integration over $p_s$ is well-defined. Thus, graphs of the type shown in Figure~\ref{fig:phi3YF1} are finite.

Problems appear, however, when the line without insertions is given by the two-point function. It turns out that then two insertions of a nonplanar tadpole, as in the graphs shown in 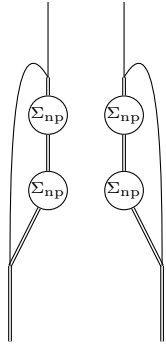
\begin{figure}[ht]
\begin{center}
\begin{tikzpicture}[scale=0.5]
 \node [circle,inner sep=0pt,draw] (s1) at (1,4) {{\tiny $\Sigma_\mathrm{np}$}};
 \node [circle,inner sep=0pt,draw] (s2) at (1,6) {{\tiny $\Sigma_\mathrm{np}$}};
 \draw[double] (0,0) to (0,2);
 \draw[double] (0,2) to (s1);
 \draw[double] (s1) to (s2);
 \draw[double] (s2) to (1,7);
 \draw (1,7) to (1,9);
 \draw (1,7) to [controls=+(120:2) and +(90:2)] (0,2);
 \node [circle,inner sep=0pt,draw] (s3) at (3,4) {{\tiny $\Sigma_\mathrm{np}$}};
 \node [circle,inner sep=0pt,draw] (s4) at (3,6) {{\tiny $\Sigma_\mathrm{np}$}};
 \draw[double] (4,0) to (4,2);
 \draw[double] (4,2) to (s3);
 \draw[double] (s3) to (s4);
 \draw[double] (s4) to (3,7);
 \draw (3,7) to (3,9);
 \draw (3,7) to [controls=+(60:2) and +(90:2)] (4,2);
\end{tikzpicture}
\end{center}
\caption{$\phi^3$ graphs in the Yang-Feldman formalism that diverge due to UV-IR mixing.}
\label{fig:phi3YF2}
\end{figure}
Figure~\ref{fig:phi3YF2}, suffice to get a divergence.
For the self-energy of these graphs, we obtain
\begin{equation}
\label{eq:phi3Sigma}
 \Sigma(k) = c_d \int \ud^dp \ \hat \Delta_1(p) \hat \Delta^3_R(k-p) \Sigma^2_\mathrm{np}(k-p) \left\{ 1 + \cos p \sigma k \right\}.
\end{equation}
Here the external momentum $k$ is confined to the upper mass shell and the momentum $p$ to the upper or the lower mass shell. In order to see how $\Sigma_\mathrm{np}(k-p)$ behaves in these two cases, we compute, for $k = (m, \V{0})$,
\begin{equation}
\label{eq:k-p}
 (\sigma (k-p^\pm))^2 = 2 \left( - p_s^2 - m^2 \pm m \sqrt{p_1^2 + p_s^2 + m^2}  \right).
\end{equation}
For $p$ on the lower mass shell, this is bounded away from zero, so that the singularity of $\Sigma_\mathrm{np}$ is not hit in \eqref{eq:phi3Sigma}. Furthermore, the rapid falloff of $\Sigma_\mathrm{np}(p)$ for $(\sigma p)^2 \to - \infty$ makes the integral well-defined in that case.

However, for $p$ on the upper mass shell, there is a $d-2$ dimensional submanifold of $\R^{d-1}$ for which $\sigma(k-p^+)$ is lightlike. To see this, note that \eqref{eq:k-p} with the $+$ sign vanishes, for a given $p_s^2$, for any $p_1$ such that
\[
 p_1^2 = \left( \left( \frac{1}{2} + \frac{p_s^2}{m^2} \right)^2 - \frac{1}{4} \right) m^2.
\]
Furthermore, we compute
\[
 \del_{p_1} (\sigma (k-p^+))^2 = \frac{2 m p_1}{\sqrt{p_1^2 + p_s^2 + m^2}},
\]
which does not vanish at the above $(p_1, p_s)$ for $p_s \neq 0$, i.e., away from the origin.  
For one nonplanar insertion that behaves as $\Sigma_\mathrm{np}(p) \sim (\sigma p)^{-2}$ for small $(\sigma p)^2$, i.e., for the nonplanar tadpole in $d=4$ or the nonplanar fish graph in $d=6$, this means that for a given $p_s$ we would have to integrate $p_1$ over a singularity of the form $(p_1 - p_1(p_s))^{-1}$. This seems like a logarithmic divergence. However, we assumed that the singularity is regularized by some $i \varepsilon$ or principal value description. But for two nonplanar insertions, we would need to define the square $\Sigma^2_\mathrm{np}(p)$, which is not well defined in the sense of H\"ormander for $(p \sigma)^2 = 0$ if the wave front set is given by \eqref{eq:wfSigma}. As the scaling degree of $\Sigma^2_\mathrm{np}$ at the submanifold $\{p| (\sigma p)^2 = 0, p \neq 0 \}$ is 2, and the submanifold has codimension 1, this square can only be renormalized at the expense of a momentum-dependent, i.e., nonlocal, counterterm, cf.~\cite{BrFr00} and Appendix~\ref{app:microlocal}. This divergence is an instance of the difficulty (i) mentioned in the introduction. As it occurs at finite momentum, it should be termed an infrared divergence. Hence, for $d=6$, already the graphs shown in Figure~\ref{fig:phi3YF2} are infrared divergent, contrary to the Euclidean case, where three insertions of nonplanar fish graphs are needed in order to make the graph infrared divergent.

\section{Conclusion}
\label{sec:Conclusion}
The aim of this work was to see whether and in what form UV-IR mixing leads to infrared divergences in the Yang-Feldman formalism. We saw that such divergences do indeed occur, due to two different mechanisms: The divergence discussed in Section~\ref{sec:phi4} was due to the fact that, for the nonplanar tadpole, $\Sigma_\mathrm{np}(p)$ is constant on the noncompact hypersurfaces $(\sigma p)^2 = \mathrm{const}$, while the problems discussed in Section~\ref{sec:phi3} stem from the integration over the singularity of $\Sigma_\mathrm{np}(p)$ in $(\sigma p)^2 = 0$.
In particular the first mechanism leads to divergences in situation that are finite in the Euclidean, e.g., in the two-dimensional case.

Thus, it seems that the introduction of nonlocal counterterms is unavoidable. As proposed in \cite{LiaoSibold}, one should try to restrict to counterterms that are functions of $(\sigma p)^2$, so that one obtains local counterterms in the commutative limit. For the Euclidean case, it was shown in \cite{Gurau} that the introduction of a $(\sigma p)^{-2}$ mass counterterm suffices to renormalize the $\phi^4_4$ model. However, the adaption of such a setting to the Lorentzian case is not straightforward. While it is easy to see that the graphs treated above can be renormalized in that way (if also a $\log \betrag{(\sigma p)^2}$ mass term is permitted), it is not clear whether this works to all orders. Some difficulties show up when one tries to tackle this problem: Because of the appearance of two propagators, the Yang-Feldman formalism is combinatorically more complicated than a treatment in terms of Feynman graphs. In particular, there is no obvious power counting. In general, a cancellation of several terms has to be taken into account to get the correct scaling. But even if there was a good notion of power counting, the introduction of terms of the form $(\sigma p)^{-2}$ would make it much more involved, as also the infrared scaling would have to be taken into account. The multiscale analysis employed in \cite{Gurau} considers the ultraviolet and the infrared regimes on the same footing. But it is far from obvious how such a multiscale analysis should look like in the Lorentzian case. It thus seems that one should try and find a way to map the Lorentzian model to a Euclidean one in order to use the powerful tools available there. However, we showed above that in the Lorentzian case we face infrared divergences even in cases that are finite in the Euclidean setting, in particular also in the two-dimensional case. Thus, a mapping between the Lorentzian and the Euclidean model, if it exists at all, must be rather nontrivial.

\subsection*{Acknowledgment}
It is a pleasure to thank Dorothea Bahns and Micha{\l} Wrochna for helpful discussions. This work was supported by the German Research Foundation (Deutsche Forschungsgemeinschaft (DFG)) through the Institutional Strategy of the University of G\"ottingen.

\appendix
\section{The wave front set and the scaling degree}
\label{app:microlocal}
We provide a short introduction to the concept of the wave front set and the scaling degree of a distribution. We recall that the singular support of a distribution is the set of points for which no open neighborhood exists on which the distribution is smooth. The wave front set generalizes this notion in that it also gives information about the direction in which the distribution diverges. To motivate the definition we recall that if $f$ is smooth and compactly supported, then its Fourier transform falls off faster than any power in momentum space, i.e., for each $N \in \N$ there is a constant $C_N$ such that
\[
 \betrag{ \hat f(k) } \leq C_N (1+\betrag{k})^{-N}.
\]
For a distribution $u$ with compact support, one defines $\Sigma(u)$ as the set of $k \in \dot \R^n = \R^n \setminus \{0\}$ for which no conic neighborhood exists in which such a bound holds. For each point $x$ one then defines
\[
 \Sigma_x (u) = \cap_{f} \Sigma (f u), \quad f \in C_0^\infty(\R^n), f(x) \neq 0.
\]
The wave front set now collects all these into a single object:
\[
 \WF(u) = \{(x,k) \in \R^n \times \dot \R^n | k \in \Sigma_x(u) \}.
\]
This notion can be lifted to any smooth manifold, where it is then interpreted as a subset of the cotangent bundle. Thus, we always interpret the second component as a cotangent vector, which means that we have to take care of the metric. 
We give the wave front set of some of the distributions that appear in this article:
\begin{align*}
 \WF(\Delta_+) & = \{ (0,k) | k^2 = 0, k^0 > 0 \} \cup \{ (x,k) | x \neq 0, x^2 = 0, k = \lambda x, k^0 > 0 \}, \\ 
 \WF(\Delta_R) & = \{ (0,k) | k \neq 0 \} \cup \{ (x,k) | x^2 = 0, x^0 > 0, k = \lambda x, \lambda \neq 0 \}, \\
 \WF(\hat \Delta_+) & = \{ (k,y) | k^2 = m^2, k_0>0, y = \lambda k, \lambda \neq 0 \}.
\end{align*}
One important feature of the wave front set is that it provides a criterion for the well-definedness of the product of two distributions, namely H\"ormander's criterion \cite[Thm.~8.2.10]{Hoermander}. It states that the product of two distributions $u,v \in \mathcal{D}'(\R^n)$ is well-defined as an element of $\mathcal{D}'(\R^n)$, provided that
\[
 \{ (x,k) \in \WF(u) | (x,-k) \in \WF(v) \} = \emptyset.
\]

Unfortunately, H\"ormander's criterion is not fulfilled for many of the products of distributions that arise in quantum field theory. The obstruction is usually (in ordinary, i.e., commutative, field theory) located at the origin, i.e., the products are well-defined on test functions that vanish in a neighborhood of the origin. The ambiguity in the extension to all test functions is governed by Steinmann's scaling degree \cite{Steinmann, BrFr00}. For $u \in \mathcal{D}'(\dot \R^n)$, i.e., a distribution on test functions vanishing in a neighborhood of the origin, it is defined as
\[
 \sd(u) = \sup \left\{ \rho \in \R | \lim_{\lambda \to \infty} \lambda^\rho \int \ud^n x \ u(\lambda x) f(x) < \infty \ \forall f \in C^\infty_0(\dot \R^n) \right\}.
\]
One can now prove that for $\sd(u) < n$ there is a unique extension $\tilde u \in \mathcal{D}'(\R^n)$ to all test functions such that $\sd(u) = \sd(\tilde u)$. For $n \leq \sd(u) < \infty$, an extension that preserves the scaling degree is still possible, but with some ambiguity: For two such extensions $u_1$ and $u_2$ we have
\[
 u_1 - u_2 = \sum_{\betrag{\alpha} \leq \sd(u)-n} c_\alpha \del^\alpha \delta.
\]
In quantum field theory, this corresponds to a finite renormalization. The concept of the scaling degree at a point was generalized by Brunetti and Fredenhagen \cite{BrFr00} to the scaling degree at a submanifold. The criterion for the existence of a unique extension to the submanifold is then that the scaling degree is less than the codimension of the submanifold.

\end{document}